\documentclass[graybox, envcountchap]{svmult}

\usepackage{mathptmx}        
\usepackage{amsmath}
\usepackage{amssymb}
\usepackage{color}
\usepackage{helvet}          
\usepackage{courier}         
\usepackage{dirtree}

\usepackage{makeidx}        
\usepackage{graphicx}        
\usepackage{subfig}

\usepackage{multicol}        
\usepackage[bottom]{footmisc}

\usepackage{hyperref}        
\hypersetup{colorlinks=true,urlcolor=blue}

\usepackage[misc]{ifsym}

\makeindex             

\begin{document}


\title{Regular Rotating Black Holes}
\author{Ram\'{o}n Torres}
\institute{(\Letter) \at Ram\'{o}n Torres, Dept. of Physics. Universitat Polit\`{e}cnica de Catalunya. Barcelona. Spain, \email{ramon.torres-herrera@upc.edu}}
%
%
\maketitle

\abstract{
The collapse of astrophysically significant bodies generates, under suitable conditions, black holes. Since one expects the generator of the black hole to be a rotating body, the black hole will also rotate. The existence of inner singularities in classical solutions for rotating black holes and the fact that General Relativity is incompatible with Quantum Mechanics lead us to seek for alternative regular models for rotating black holes. The interest in singularity-free rotating black holes has grown significantly in recent years, as shown by the increase in the number of published papers devoted to it.
Undoubtedly, the latest observational developments (LIGO-VIRGO-KAGRA collaborations, the Event Horizon Telescope or, in the near future, the LISA project) and the possibility to probe our theoretical predictions have greatly contributed to awaken the interest.\\
This text discusses the general characteristics of regular rotating black holes. These include the conditions needed to guarantee the absence of singularities and the consequences that such conditions entail for the violation of the energy conditions in black hole models. It is argued that regular rotating black holes do not require an extension through their inner disk, contrary to classical rotating black holes. In this way, the problems with negative-mass interpretations and causality violations appearing in classical solutions could be avoided. It has been included a discussion on the maximal extension and the usual global causal structure expected for these spacetimes. The different methods for obtaining regular rotating black holes are treated, including the use of the generalized Newman-Janis algorithm as an alibi to derive regular rotating black holes from regular spherically symmetric static ones. The text also provides an introduction to the thermodynamics and phenomenology of rotating black holes.}


\section{A glance at classical rotating black holes}
\label{secKerr}

Most astrophysically significant
bodies are rotating. If a rotating body collapses, the rate of rotation
will speed up, maintaining constant angular momentum. Through a rather complicated process, the body could finally generate a black hole which would be a \emph{rotating black hole} (\emph{RBH}). From a classical point of view (\textit{no-hair conjecture} \cite{gravit}) the resulting spacetime will be described by a Kerr solution (or a Kerr-Newman solution, in the charged case).
This implies that an idealized classical model of a lonely rotating body eventually generates an axially symmetric, stationary and asymptotically flat spacetime with certain horizons, a specific causal structure and a curvature singularity.

In order to compare these characteristics with those of regular RBH, let us now briefly summarize them for the classical uncharged RBH solution. (The reader can consult, for example, \cite{Griff}\cite{gravit} and references therein for more information). In Boyer-Lidquist (B-L) coordinates $\{t,r,\theta,\phi\}$, the Kerr metric takes the form
\begin{equation}\label{gKerr}
ds^2=-\frac{\Delta}{\Sigma} (dt-a \sin^2\theta d\phi)^2+
\frac{\Sigma}{\Delta} dr^2+\Sigma d\theta^2+\frac{\sin^2\theta}{\Sigma}(a dt-(r^2+a^2)d\phi)^2,
\end{equation}
where
\[
\Sigma=r^2+a^2 \cos^2\theta, \hspace{1cm} \Delta=r^2-2 m r+a^2,
\]
$m$ is the black hole mass\footnote{In case the RBH is also charged, then it is described by using the Kerr-Newman solution in which $m$ should be replaced by $m-e^2/(2r)$, where $e$ is the total charge of the RBH.} and $a$ is a \textit{rotation parameter} that measures the (Komar) angular momentum per unit of mass \cite{gravit}. The spacetime is type D whenever $m\neq 0$.

If $m\neq 0$ there is a curvature singularity at $(r=0, \ \theta=\pi/2)$, as can be shown by the divergence of the curvature invariant $R_{\alpha\beta\gamma\delta} R^{\alpha\beta\gamma\delta}$.
Remarkably, for $a\neq 0$ and $\theta\neq \pi/2$, a surface defined by $t=$constant and $r=0$ is singularity-free and has metric
\[
ds_2^2=a^2 \cos^2 \theta d\theta^2+a^2 \sin^2 \theta d\phi^2=dx^2+dy^2,
\]
where the coordinate change $x\equiv a \sin\theta \cos \phi$, $y\equiv a \sin\theta\sin\phi$ has been made to make explicit that the surface is flat.
The curvature singularity corresponds to the \textit{ring} $x^2+y^2=a^2$, while the flat surface corresponds to $x^2+y^2<a^2$. As it is customary, we will call this flat surface \textit{the disk}.

In this way, the curves that reach $r=0$ with $\theta\neq \pi/2$ are reaching a regular point in the disk. In order to continue the curves it is usually argued that an analytic extension of the spacetime has to be obtained through $r=0$. The procedure requires letting the coordinate $r$ to take negative values \cite{H&E}. The $r<0$ extended spacetime can be seen as a negative mass spacetime. Causality violations occur in the extended spacetime \cite{Carter1968a}.

The metric has a coordinate singularity at $\Delta=0$, which can be easily removed by a coordinate change \cite{B&L}. At $\Delta=0$ the hypersurface $r=$constant becomes light-like and no observer can remain at the specific value for $r$, thus the hypersurface is called a \textit{null horizon}. In the Kerr case, if $m^2>a^2$ there are two roots: $r^{Kerr}_\pm=m\pm \sqrt{m^2-a^2}$, where the null horizon $r^{Kerr}_+$ is an event horizon, while the null horizon $r^{Kerr}_-$ is a Cauchy horizon. The limiting case $m^2=a^2$ has a degenerate null horizon and the spacetime is called the \textit{extreme} Kerr black hole. If  $m^2<a^2$ there are no roots for $\Delta=0$ and the curvature singularity is naked. This is the so called \textit{hyperextreme} case.

\section{Kerr-like Rotating Black Holes}
\label{KRBHs}

Several authors have suggested that the existence of singularities in the solutions of General Relativity has to be considered as a weakness of the theory rather than as a real physical prediction.
The problem of obtaining singularity-free models for black holes was first approached for spherically symmetric black holes. In this context, some authors introduced non-standard energy-momentum tensors mainly acting in the core of the black hole (see, for example, \cite{A-BI}\cite{A-BII}\cite{B&V}\cite{Bardeen}). However, most authors expect that the inclusion of quantum theory in the description of black holes could avoid the existence of their singularities (see, for example, \cite{A&B2005}\cite{B&R}\cite{Frolov2014}\cite{G&P2014}\cite{Hay2006}\cite{H&R2014}\cite{dust2014} and references therein).

We do not yet have a mature and reliable candidate for a quantum theory of gravity so that it is difficult to accurately describe even non-rotating (quantum) black holes. On the other hand, a glimpse back into classical black hole history indicates that finding an accurate description of a (quantum) rotating black hole could be even much more difficult: Kerr solution was only discovered following 48 years of struggle after the Einstein field equations were first developed. There are some works on approximated solutions, only valid in the slow-rotation limit \cite{P&C}\cite{Y&Y}. Unfortunately, they are not adequate enough to be used in astrophysical observations.
In this way, it is necessary to try phenomenological approaches to check for possible models of regular RBHs and their implications, including the possibility of observable astrophysical predictions.

Even if a regular RBH model comes from an approach to Quantum Gravity Theory, we will assume in this chapter that it can be reasonably well described by a manifold endowed with its corresponding metric. Nevertheless, it should be taken into account that, in the absence of a full Quantum Gravity Theory, probably one can only guarantee this to be a good description of the RBH up to the high curvature planckian regime.

Recently, there have appeared different proposals
for \emph{regular} rotating black holes spacetimes with their corresponding metrics (see section \ref{secObt}).
While they have been obtained by different approaches, most of them share a common \textit{Kerr-like form}.
The general metric corresponding to this kind of RBH, was found by G\"{u}rses-G\"{u}rsey \cite{GG} as a particular rotating case of the algebraically special Kerr-Schild metric:
\begin{equation}\label{GGg}
ds^2=(\eta_{\alpha \beta} +2 H k_\alpha k_\beta) dx^\alpha dx^\beta,
\end{equation}
where $\eta$ is the metric of Minkowski, $H$ is a scalar function and $\vec k$ is a light-like vector both with respect to the spacetime metric and to Minkowski's metric.
Specifically, in Kerr-Schild coordinates $\{\tilde{t},x,y,z\}$ the  G\"{u}rses-G\"{u}rsey metric (\ref{GGg}) corresponds with the choices
\[
H=\frac{\mathcal M (r) r^3}{r^4+a^2 z^2}
\]
and
\[
k_\alpha dx^\alpha =-\frac{r (x dx+y dy) -a (x dy-y dx)}{r^2+a^2}-\frac{z dz}{r}-d\tilde{t},
\]
where $r$ is a function of the Kerr-Schild coordinates implicitly defined by
\begin{equation}\label{defr}
r^4-r^2 (x^2+y^2+z^2-a^2) -a^2 z^2 =0,
\end{equation}
$\mathcal M (r) $ is known as the \textit{mass function}
and the constant $a$ is a rotation parameter.

This metric can be written in Boyer-Lindquist-like coordinates by using the coordinate change defined by
\begin{eqnarray*}
x+i y&=&(r+i a) \sin\theta\exp\left[i\int(d\phi+\frac{a}{\Delta}dr)\right]\\
z&=&r \cos\theta\\
\tilde{t}&=&t+\int \frac{r^2+a^2}{\Delta} dr-r.
\end{eqnarray*}
where now $\Delta=r^2-2 \mathcal M(r) r+a^2$.
The resulting metric takes the form
\begin{equation}\label{gIKerr}
ds^2=-\frac{\Delta}{\Sigma} (dt-a \sin^2\theta d\phi)^2+
\frac{\Sigma}{\Delta} dr^2+\Sigma d\theta^2+\frac{\sin^2\theta}{\Sigma}(a dt-(r^2+a^2)d\phi)^2,
\end{equation}
where, again, $\Sigma=r^2+a^2 \cos^2\theta$.
Note that this metric reduces to Kerr's solution in B-L coordinates if $\mathcal M(r)=m$=constant and that it reduces to the (charged) Kerr-Newman solution if $\mathcal M(r)=m-e^2/(2r)$, where $e$ is the charge.

As in Kerr's solution, for $\theta\neq\pi/2$, the surface $t=$constant and $r=0$ is a flat surface (corresponding to $x^2+y^2<a^2$) that will be called \textit{the disk}. Likewise, \textit{the ring} corresponds to $\theta=\pi/2$ (or $x^2+y^2=a^2$).

In order to analyze the general properties of the RBH spacetime
we will use the following null tetrad-frame:
\begin{eqnarray*}
\mathbf{l} &=&\frac{1}{\Delta} \left( (r^2+a^2) \frac{\partial}{\partial t}+\Delta \frac{\partial}{\partial r}+a \frac{\partial}{\partial \phi}\right),\\
\mathbf n &=&\frac{1}{2 \Sigma} \left( (r^2+a^2) \frac{\partial}{\partial t}-\Delta \frac{\partial}{\partial r}+a \frac{\partial}{\partial \phi}\right),\\
\mathbf m &=& \frac{1}{ \sqrt{2} \varrho } \left(i a \sin\theta \frac{\partial}{\partial t}+\frac{\partial}{\partial \theta}+i \csc\theta \frac{\partial}{\partial \phi} \right),\\
\mathbf{\bar m} &=& \frac{1}{ \sqrt{2} \bar\varrho } \left(-i a \sin\theta \frac{\partial}{\partial t}+\frac{\partial}{\partial \theta}-i \csc\theta \frac{\partial}{\partial \phi} \right),
\end{eqnarray*}
where $\varrho\equiv r+i a \cos\theta$, $\bar\varrho\equiv r-i a \cos\theta$ and the tetrad is normalized as follows $\mathbf l^2=\mathbf n^2=\mathbf m^2=\mathbf{\bar m}^2=0$ and $\mathbf l\cdot \mathbf n=-1= -\mathbf{m}\cdot \mathbf{\bar m}$.

\begin{theorem}\label{PTD}\cite{TorresReg}
The RBH metric (\ref{gIKerr}) with $\mathcal M(r)\neq 0$ is Petrov type D and the two double principal null directions are $\mathbf l$ and $\mathbf n$.
\end{theorem}

We can also define a real orthonormal basis $\{\mathbf{t}, \mathbf{x}, \mathbf{y}, \mathbf{z } \}$
formed by a timelike vector $\mathbf{t}\equiv (\mathbf{l}+\mathbf{n})/\sqrt{2}$ and three spacelike vectors: $\mathbf{z}\equiv (\mathbf{l}-\mathbf{n})/\sqrt{2}$,  $\mathbf x=(\mathbf m +\bar{\mathbf m})/\sqrt{2}$ and $\mathbf y=(\mathbf m -\bar{\mathbf m}) i/\sqrt{2}$. Then,
$\mathbf t$ and $\mathbf z$ are two eigenvectors of the Ricci tensor with eigenvalue \cite{TorresReg}
\begin{equation}\label{lambda1}
\lambda_1=\frac{2 a^2 \cos^2{\theta} \mathcal M'+r \Sigma \mathcal M''}{\Sigma^2}.
\end{equation}
$\mathbf x$ and $\mathbf y$ are two eigenvectors of the Ricci tensor with eigenvalue
\begin{equation}\label{lambda2}
\lambda_2=\frac{2 r^2 \mathcal M'}{\Sigma^2}.
\end{equation}

In this way, the Ricci tensor can be written as
\begin{equation}\label{Ricci}
R_{\mu\nu}= \lambda_1\, (-t_\mu t_\nu+z_\mu z_\nu)+ \lambda_2 (x_\mu x_\nu+y_\mu y_\nu),
\end{equation}

what shows the following

\begin{theorem}\cite{TorresReg}
The metric (\ref{gIKerr}) with $\mathcal M\neq$constant is Segre type [(1,1) (1 1)].
\end{theorem}

 Note that the $\mathcal M\neq$constant case is precisely the case we are interested in for our regular RBHs, since the $\mathcal M=$constant (i.e., Kerr's case) is singular.

\section{Regularity in Kerr-like Rotating Black Holes }\label{secRegu}

In order for the model of a RBH to be regular it should be devoid of curvature singularities. Let us now specifically analyze the absence of \emph{scalar} curvature singularities. We say that there is a \textit{scalar curvature singularity}
in the spacetime if any scalar invariant polynomial in the Riemann tensor diverges when approaching it along any incomplete curve.

It is well-known \cite{Weinberg} that an arbitrary spacetime possesses at most 14 second order algebraically independent invariants. The finiteness of \emph{all} the invariants is a necessary and sufficient condition for the absence of scalar curvature singularities.

A minimum set of reliable independent invariants for the RBH spacetime exists. This can be shown thanks to the following result by Zakhary and McIntosh \cite{ZM}

\begin{theorem}
The algebraically complete set of second order invariants for a Petrov type D spacetime and Segre type [(1,1) (1 1)] is $\{\mathcal R,I,I_6,K\}$.
\end{theorem}
Apart form the well-known curvature scalar $\mathcal R$, the rest of the invariants are defined as\footnote{Here the invariants are written in tensorial form. See \cite{ZM} for their spinorial form.}
\begin{eqnarray*}
I_6&\equiv&
\frac{1}{12}
{S_\alpha}^\beta {S_\beta}^\alpha,\\
I &\equiv&
\frac{1}{24}
\bar{C}_{\alpha\beta\gamma\delta}
\bar{C}^{\alpha\beta\gamma\delta},\\
K &\equiv&
\frac{1}{4}
\bar{C}_{\alpha\gamma\delta\beta}
S^{\gamma\delta} S^{\alpha\beta},
\end{eqnarray*}
where ${S_\alpha}^\beta \equiv {R_\alpha}^\beta-
{\delta_\alpha}^\beta \mathcal{R}/4$ and
$\bar{C}_{\alpha\beta\gamma\delta}\equiv
(C_{\alpha\beta\gamma\delta} +i\ *C_{\alpha\beta\gamma\delta})/2$
is the complex conjugate of the selfdual Weyl tensor
being $*C_{\alpha\beta\gamma\delta}\equiv
\epsilon_{\alpha\beta\mu\nu} C^{\mu\nu}_{\ \ \gamma\delta}/2$ the dual of the Weyl tensor.
Note that $\mathcal R$ and $I_6$ are real, while $I$ and $K$ are complex. Therefore for this type of spacetimes there are only 6 independent real scalars.

It trivially follows from our previous propositions
\begin{corollary}\cite{TorresReg}
The algebraically complete set of second order invariants for the RBH metric (\ref{gIKerr}) is $\{\mathcal R,I,I_6,K\}$.
\end{corollary}

Similarly to Kerr's case, a straightforward inspection of the metric (\ref{gIKerr}) tell us that it is singular if there are values of $r$ such that $\Delta=0$ and if $\Sigma=0$. However, $\Delta=0$ is not a scalar curvature singularity since the curvature scalars do not diverge there.
for the values of $r$ ($\neq 0$) where $\Delta=0$.
It is simply a coordinate singularity that can be removed through a coordinate change. (See section \ref{Horizons}).

Scalar curvature singularities do may appear if $\Sigma=0$ or, in other words, in $(r=0,\theta=\pi/2)$. (We already confirmed this possibility in section \ref{secKerr} for the particular case of Kerr's solution).
Now, by explicitly computing the complete set of scalars in our case, one directly gets a necessary and sufficient condition for the absence of scalar curvature singularities:

\begin{theorem}\label{teorema}\cite{TorresReg}
Assuming a RBH metric (\ref{gIKerr}) possessing a $C^3$ function $\mathcal M(r)$, all its second order curvature invariants will be finite at $(r=0,\theta=\pi/2)$ if, and only if,
\begin{equation}\label{condisreg}
 \mathcal M (0)= \mathcal M' (0)= \mathcal M'' (0)=0 .
\end{equation}
\end{theorem}

The absence of curvature singularities is a necessary condition in order to have a regular rotating black hole. The theorem allows to control the specific case of \emph{scalar curvature singularities}, which arguably are the most serious type of curvature singularities. However, since scalar polynomials do not fully characterize the Riemann tensor, it does not cover the possibility of the existence of curvature singularities with respect to a parallelly propagated basis (\textit{p.p. curvature singularity})\cite{H&E}. This possibility has not yet been fully analyzed in the literature.

\section{Violation of the energy conditions}\label{ecs}

The energy conditions were first developed in the framework of Einstein's General Relativity. These are conditions imposed on the energy-momentum tensor of the spacetime as a means of ensuring plausible matter-energy contents \cite{H&E}. Even if here we are not confined to General Relativity we can take profit of the energy conditions by considering the existence of an \textit{effective energy-momentum tensor} defined through
\[
T_{\mu\nu}\equiv R_{\mu\nu}-\frac{1}{2} \mathcal R g_{\mu\nu}.
\]
In our more general context, it is usually argued that it seems reasonable to demand the spacetime describing a realistic isolated RBH to fulfill the standard energy conditions in  asymptotically flat regions (thus, imitating the classical RBH solutions at large distances/low curvatures). Nevertheless, probably it would be more accurate to say that one should expect extremely small violations of the energy conditions in the asymptotically flat regions. This is due to the fact that, as pointed out by Donoghue \cite{Dono}, the standard perturbative quantization of Einstein gravity leads to a well-defined, finite prediction for the leading large distance correction to Newton’s potential. Specifically, it is shown that quantum effects produce deviations in the gravitational field for spherically symmetric fields of the order $G m l_p^2/r^3$ whenever $r\gg 2 m$, being $l_p$ Planck's length. This implies an extremely small reduction of the classically expected (negative) gravitational field. In the weak field approximation this leads to an effective energy-momentum tensor that (\textit{slightly}) violates the dominant energy conditions \cite{B&R}\cite{TorresVoids}.

Let us now treat the behaviour of the energy conditions in the region around $r=0$ for regular RBH. If we take the expression obtained for the Ricci tensor (\ref{Ricci}) one can explicit $\mathbf{T}$ for a RBH as
\[
T_{\mu\nu}=-\lambda_2 (-t_\mu t_\nu + z_\mu z_\nu)- \lambda_1 (x_\mu x_\nu+ y_\mu y_\nu).
\]
Since $\mathbf{T}$ diagonalizes in the orthonormal basis $\{\mathbf{t}, \mathbf{x}, \mathbf{y}, \mathbf{z } \}$, the RBH spacetime possesses an (effective) energy-momentum tensor of type I \cite{H&E}. The (effective) density being $\mu=\lambda_2$ and the (effective) pressures being $p_x=p_y=-\lambda_1$ and $p_z=-\lambda_2$. The weak energy conditions \cite{H&E} require $\mu\geq 0$ and $\mu+p_i\geq 0$. In other words, in this case they require
\[
\lambda_2 \geq 0 \hspace{1 cm} \mbox{and} \hspace{1 cm} \lambda_2-\lambda_1\geq 0.
\]
By using this and expressions (\ref{lambda1}) and (\ref{lambda2}) it is easy to show the following
\begin{theorem}\cite{TorresReg}
Assume that a \emph{regular} RBH has a function $\mathcal M (r)$ that can be approximated by a Taylor polynomial around $r=0$, then the weak energy conditions should be violated around $r=0$.
\end{theorem}

Note that for this type I effective energy-momentum the violation of the weak energy condition also implies the violation of the \textit{dominant} and the \textit{strong} energy conditions.

In this way, no model with \textit{normal} matter (matter satisfying the energy conditions) can produce a \emph{regular} rotating black hole of the type (\ref{gIKerr}).
However, the violation of the WEC around $r=0$ is not problematic since it is well-known that quantum effects can violate the WEC (Casimir effect). Moreover, singularity theorems require the spacetime to fulfill some energy condition in order to predict the existence of singularities. In this sense, the violation of energy conditions just \textit{helps} to avoid the existence of singularities \footnote{Of course, regularity can also be obtained by violating other assumptions in the singularity theorems}.

\section{Extensions beyond the disk}\label{secr0}

As stated in section \ref{secKerr}, for Kerr's solution
one could considered the possibility of extending the spacetime through the disk.
Now, in order to analyze the general situation for regular RBHs with metric (\ref{GGg}), let us proceed with an analysis similar to the one usually carried out for the classical RBH case. Consider the metric component
\begin{equation}\label{gtt}
g_{tt}=-1+ \frac{2 \mathcal M (r) r^3}{r^4+a^2 z^2}.
\end{equation}
Let us imagine and observer crossing $r=0$ moving in the $z$ \textit{axis} ($x=y=0$).
If we persist in considering $r$ as non-negative, then (\ref{defr}) implies that $r=|z|$ along the trajectory of the observer, so that along it
\begin{equation}\label{gttz}
g_{tt}=-1+ \frac{2 \mathcal M (|z|) |z|}{z^2+a^2}.
\end{equation}
The numerator in the fraction indicates that the derivative of this metric component along the axis, as well as the Christoffel symbols and the extrinsic curvature of the surface can be discontinuous across the disk depending on the chosen mass function. In particular, this discontinuity occurs if the mass function is constant (Kerr's case).

\begin{figure}[ht]
\includegraphics[scale=0.7]{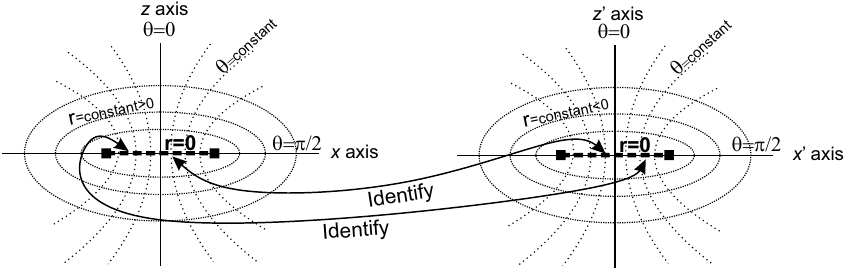}
\caption{\label{extensionr0} In Kerr's case and approach a), the extension through $r=0$ is obtained by identifying the top of the surface ($r=0, t=$constant) in the hypersurface described by coordinates $\{x,y,z\}$ with the bottom of the surface ($r=0, t=$constant) in the hyersurface described by coordinates $\{x',y',z'\}$, and vice versa. Only the $y=0$, $y'=0$ sections of theses hypersurfaces are represented here.}
\end{figure}

Historically, the differentiability problems in Kerr's RBH have been approached:
\begin{itemize}
\item[a)] By analytically extending the spacetime through $r=0$ with negative values for $r$. This requires considering two spacetimes, one with positive $r$ and another with negative $r$ and properly identifying points in their $r=0$ surfaces by a standard procedure which is ilustrated in figure \ref{extensionr0} (see, for example, \cite{H&E})\footnote{Note also that this approach has been criticized in \cite{GC&M}.}.
\item[b)] By considering the discontinuity of the derivatives of the metric components in the disk and, thus, the discontinuities in the second fundamental form, as indicating the presence of a thin shell in the surface \cite{IsraelThin}.
\end{itemize}

It was soon noticed that by following approach a), and extending through $r=0$ with negative values of $r$, can produce closed causal curves and, thus, causality problems \cite{H&E}. This will be treated, for the general case of RBH, in section \ref{secCaus},

With regard to approach b), it not only leads to the concentration of mass-energy in a infinitesimally thin surface, but it also needs its matter to move faster than light\cite{Israel}\cite{Hamity}.

\begin{figure}[ht]
\includegraphics[scale=.7]{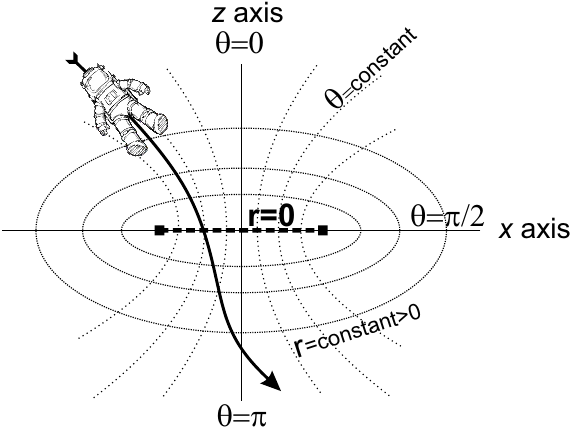}
\caption{\label{caseB} For regular RBH no extension through $r=0$ is required. An observer crossing the surface ($r=0,t=$constant) from positive $z$ to negative $z$ following a non-geodesic time-like curve can stay in its original spacetime. Along the trajectory of the observer $r$ just decreases until reaching the surface $r=0$ (disk), where it increases again.}
\end{figure}

At first sight, the situation for general regular RBH looks much better. Assuming that the regular RBH has a mass function $\mathcal M (r) \sim r^n$ with $n\geq 3$ around $r=0$, the metric component along the trajectory (\ref{gttz}) will not have differentiability problems in $z=0$ ($\partial_z g_{tt}(z=0)=0$) (and, in fact, it will be at least $C^n$). This suggests that the extension through the disk could not be necessary for regular RBH.
In order to show it, one has to go beyond a particular trajectory intersecting the disk and beyond the analysis of a single metric component. Let's start by noticing that while approaching a point in the disk ($x^2+y^2<a^2$), according to (\ref{defr}), the function $r$ approaches zero whenever $z$ approaches zero and vice versa. If we insist in having a positive $r$, we get, solving for $r$ in (\ref{defr}), that around $z=0$
\[
r \simeq \frac{|a|}{\sqrt{a^2-(x^2+y^2)}} |z|
\]
If we introduce this into the metric component (\ref{gtt}) and considering a mass function $\mathcal M (r) \sim r^n$ with $n\geq 3$ around $r=0$, we see that the metric component takes the form
\[
g_{tt} \simeq -1+\frac{f(x,y) |z|^{n+1}}{g(x,y) z^2+a^2},
\]
where $f$ and $g$ are finite differentiable functions in the disk. In this way, $g_{tt}$ is differentiable at the disk. (In particular, again ($\partial_z g_{tt}(z=0)=0$)).
The reader can check that the same situation is found for the rest of metric components. Let us only remark that the metric will not be analytic at the disk. Not all metric components will be infinitely differentiable. For example, even if the particular metric component (\ref{gtt}) and for odd $n$ is $C^\infty$ other metric components like
\[
g_{tz}=\frac{2\mathcal{ M}(r) r^2 z (a y+x r)}{(a^2+r^2)(a^2 z^2+r^4)}
\]
are not. Nevertheless, such a degree of differentiability is not required at all\footnote{Usually the metric is required to be at least $C^2$ \cite{H&E}. However, many authors consider this degree of differentiability too restrictive.}. In this way, regular RBH do not have differentiability problems and an extension through the $r=0$ is not needed\footnote{Let us comment that, even if not mathematically needed, the possibility of extending through $r=0$ with negative values of $r$ exists, in principle, for all regular RBH.}. An observer could cross through $r=0$ while remaining in the ($r\geq 0$) spacetime. (See figure \ref{caseB})). Furthermore, most of the problems found in Kerr's RBH would be nonexistent.

\section{Maximal extensions, null horizons and global structure}\label{Horizons}

Metric (\ref{gIKerr}) in Boyer-Lindquist-like coordinates has a coordinate singularity at $\Delta=0$ that can be eliminated through a coordinate change in order to obtain the maximally extended spacetime. The procedure is similar to the one usually carried out in Kerr's solution \cite{B&L}. For example, one can perform a coordinate change from B-L-like coordinates $\{t,r,\theta,\phi\}$ to advanced Eddington-Finkelstein-like or \textit{Kerr-like coordinates} $\{u,r,\theta,\varphi\}$, where $u$ is a light-like coordinate, through \footnote{By means of these kind of coordinate changes -advanced and retarded- the \emph{maximal} extension is obtained by following the procedure in \cite{B&L}.}
\begin{equation}\label{uphi}
u\equiv t+ \int \frac{r^2+a^2}{\Delta} dr \hspace{1cm};\hspace{1cm} \varphi=\phi+\int \frac{a}{\Delta} dr.
\end{equation}

In these Kerr-like coordinates the metric takes the form
\begin{eqnarray*}
ds^2&=&-\left(1-\frac{2 \mathcal M(r) r}{\Sigma} \right) du^2+2 dudr+\Sigma d\theta^2-2a \sin^2\theta dr d\varphi\\
&+&\left( r^2+a^2+\frac{2 a^2 \mathcal M(r) r  \sin^2\theta}{\Sigma} \right) d\varphi^2-\frac{4 \mathcal M (r) r a}{\Sigma} \sin^2\theta du d\varphi
\end{eqnarray*}
and the problems with $\Delta=0$ disappear.

The causal character of the $r=$constant hypersurfaces is defined by the sign of $g^{rr}=\Delta/\Sigma$. Since $\Sigma>0$ (except at the ring $r=0,\theta=\pi/2$), the causal character of the $r=$constant hypersurfaces depends on the sign of $\Delta$.
In particular, this hypersurface will be light-like if $\Delta=0$, so that observers will not be able to remain at $r=$ constant at these particular hypersurfaces, thus called \textit{null horizons}.
We have already treated the null horizons in Kerr's solution in section \ref{secKerr}.
Now, in order to get the null horizons in the general RBH case we should solve
\begin{equation}\label{eqdelta}
\Delta=r^2-2 \mathcal M(r) r+a^2=0.
\end{equation}
Without the knowledge of a specific $\mathcal M(r)$ it is not possible to know the \emph{exact} position of the horizons. Nevertheless, one can analyze the general behaviour of the horizons by taking into account the following considerations:
\begin{itemize}
\item If we assume an asymptotically flat spacetime, at large distances $\mathcal M(r)\simeq m=$constant, so that one (approximately) recovers the behaviour for the Kerr solution. Then $\Delta>0$ and $r$ will be a spacelike coordinate.
\item For $r\simeq 0$ ($a\neq 0$) a regular RBH has $\Delta>0$ thanks to the effect of the rotation and, again, $r$ will be a spacelike coordinate. (Note that this already happens in the classical Kerr solution).
\item If we assume the existence of a RBH and, thus, the existence of an exterior horizon $r_+$ (solution of $\Delta=0$) then the continuity of $\Delta$ and the two previous items imply either a single horizon (\textit{extreme} RBH), two horizons $r_-$ and $r_+\ (>r_-) $ or, in general, an even number of horizons.
\item If no solutions of (\ref{eqdelta}) exist, then no null horizons exist and we are in a \textit{hyperextreme} case. The regular rotating astrophysical object without an event horizon is not properly a black hole. The regularity implies that, contrary to the classical case, there is not a naked singularity.
\end{itemize}

In practice, the usual regular RBH in the literature has one or two null horizons, as in the classical case. This is not surprising if one considers deviations from General Relativity as coming from Quantum Gravity effects. Then, based on a simple dimensional analysis, one could expect the Planck scale to be the most natural scale in which to expect the departure from General Relativity to occur, what would imply only strong deviations from the classical solution around $r\sim r_{Planck}$ and, thus, only small corrections to the horizons (at least for RBH with masses much larger than the planckian mass). One also expects that associated with non-singular RBH there would be a \textit{weakening of gravity}. An effect which should be very important at high curvature scales. In this way, comparing with the classical case, it is usual to obtain bigger inner horizons and smaller outer horizons.
Of course, the Planck scale approach could turn out to be too naive and bigger deviations from the classical solutions could be possible, what would be good news for the observational aspects of RBH (see section \ref{secPheno}).

\begin{figure}[htp]
\includegraphics[scale=.7]{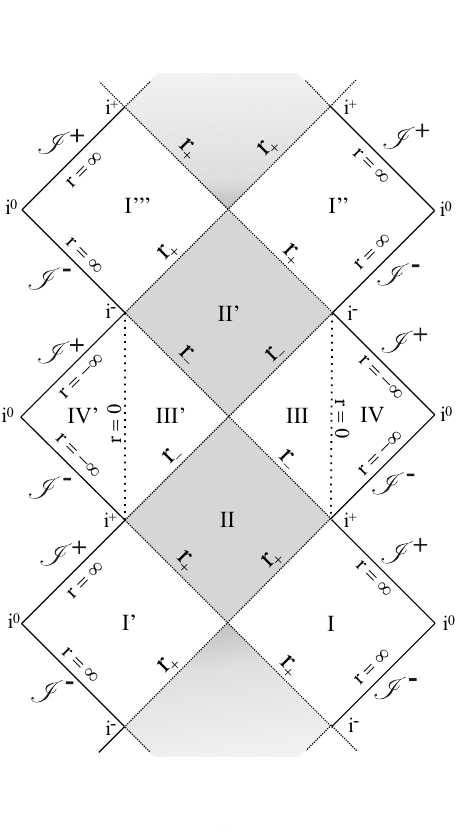}
\caption{\label{Pbiggera} Penrose diagram for Kerr's RBH with two horizons. The spacetime has been extended through $r=0$ to asymptotically flat regions with negative values for $r$ (IV or IV'). The grey regions are the regions where the coordinate $r$ is timelike. Starting from the asymptotically flat region I, one could enter region II by traversing the \emph{event horizon} $r_+$. Region III could next be reached by traversing the \emph{Cauchy horizon} $r_-$. Then, the asymptotically flat region IV could be reached by passing through the regular $r=0$. Note that the diagram is valid for $\theta\neq \pi/2$. The diagram with $\theta= \pi/2$ will require to draw the ring singularity.}
\end{figure}

\begin{figure}[htp]
\includegraphics[scale=.7]{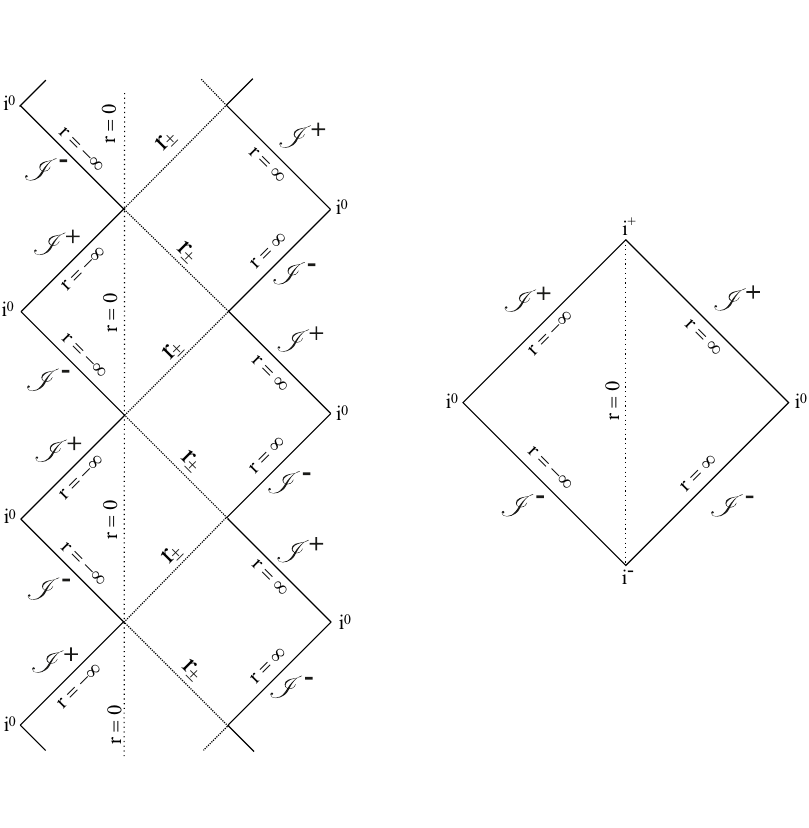}
\caption{\label{PHE} Penrose diagrams for Kerr's extreme rotating black hole (to the left) and for the hyperextreme case (to the right). In the extreme case there is only one horizon denoted by $r_\pm$ in which the coordinate $r$ is lightlike. $r$ is never timelike. $r_\pm$ acts both as an event and as a Cauchy horizon.
 In the hyperextreme case there are no horizons and $r$ is always spacelike. In both cases, the spacetime has been extended through $r=0$ to an asymptotically flat region with negative values for $r$. (Note that, again, the diagrams are valid for $\theta\neq \pi/2$). The diagram with $\theta= \pi/2$ will require to draw the ring singularity.}
\end{figure}

Nevertheless, in order to illustrate the global causal structure of regular RBH let us follow the approach of small perturbations with respect to the classical horizons.
We will compare this regular RBH causal structure with the usual one for Kerr's RBH, where we extend the spacetime through $r=0$ into negative values for $r$. There are three possible qualitatively different causal structures for Kerr's RBH spacetime which are represented in the Penrose diagrams of figure \ref{Pbiggera} (for the case with two null horizons) and of figure \ref{PHE} (for the \textit{extreme} case and the \textit{hyperextreme} case).

\begin{figure}[htp]
\includegraphics[scale=.7]{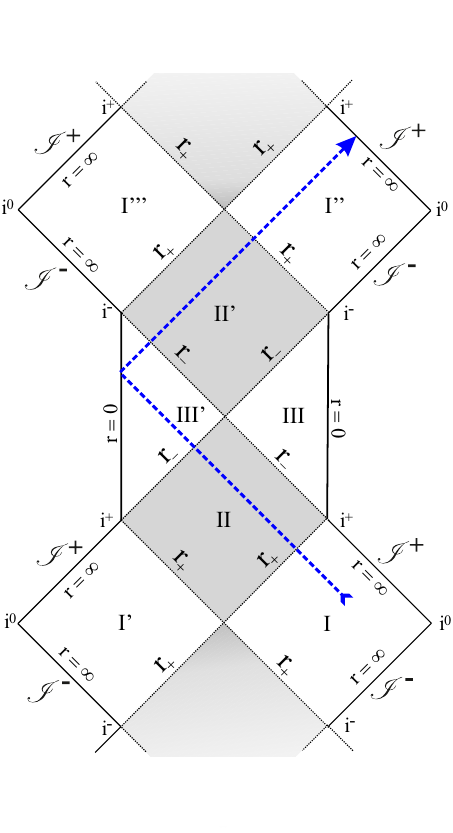}
\caption{\label{Pbiggera3} Penrose diagram for a regular rotating black hole with two null horizons. In this case an extension through $r=0$ is not required. The grey regions are the regions where the coordinate $r$ is timelike. We have depicted a light-like geodesic (dashed blue line) that, starting from the asymptotically flat region I, enters region II by traversing the \emph{event horizon} $r_+$. Then it reaches region III' by traversing the null horizon $r_-$. The value of $r$ first decreases along the geodesic until reaching $r=0$, where it increases again. It makes it to another null horizon $r_-$, enters region II', traverses another event horizon $r_+$ to enter the asymptotically flat region I'' where it travels towards the future null infinity. (Note that, since there are not singularities, the diagram is valid for all $\theta$). }
\end{figure}

\begin{figure}[htp]
\includegraphics[scale=.7]{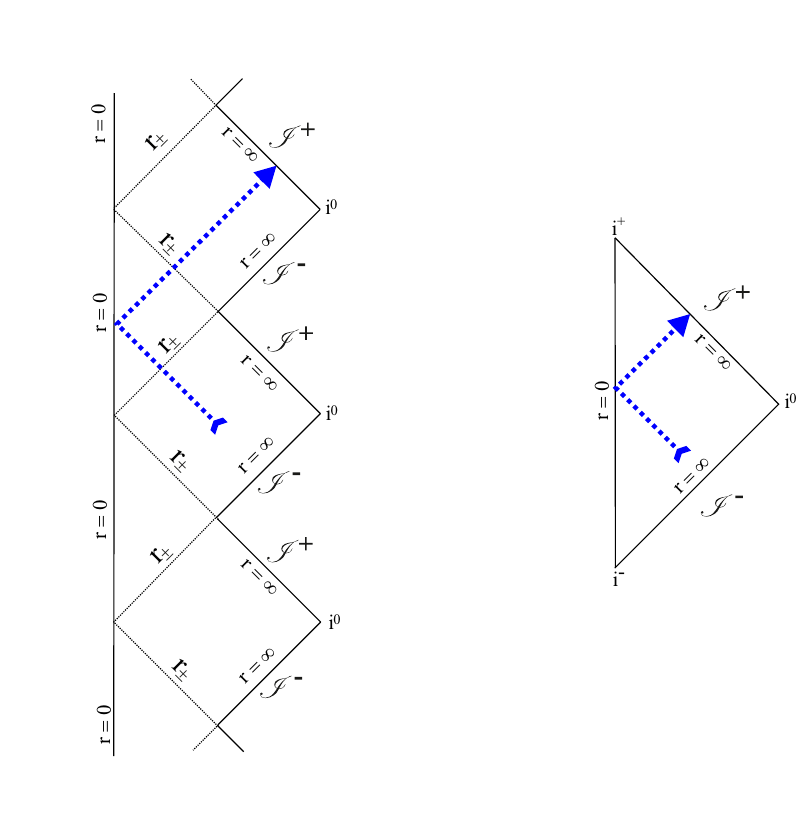}
\caption{\label{PHE3} Penrose diagrams for an extreme regular rotating black hole (to the left) and for a hyperextreme case (to the right). In both cases, an extension through $r=0$ is not required. In the extreme case there is only one horizon denoted by $r_\pm$ in which the coordinate $r$ is lightlike. $r$ is never timelike. $r_\pm$ acts as an event horizon.
 In the hyperextreme case there are no horizons and $r$ is always spacelike. In both diagrams a light-like geodesic (dashed blue line) travels towards $r=0$, traverses the disk (or ring) and continues its travel towards the future null infinity.  (Note that, again, since there are not singularities, the diagrams are valid for all $\theta$).}
\end{figure}

If we are in the regular RBH case then there is no need for an extension through $r=0$. We can have three possible qualitatively different causal structures for the BH spacetime which are represented in the Penrose diagrams of figure \ref{Pbiggera3} (for the case with two null horizons) and of figure \ref{PHE3} (for the \textit{extreme} case and the \textit{hyperextreme} case).

The absence of an event horizon in the hyperextreme case is interesting, since this implies that an observer could receive information from the inner high curvature regions near $r=0$. In principle, this could be used to observationally test the different approaches to Quantum Gravity.
The problem is whether such RBHs are feasible. In the framework of General Relativity, it does not seem possible to obtain such high speed RBH ($a^2>m^2$) from a collapsing star and any attempt to overspin an existing black hole destroying its event horizon has failed, in agreement with the weak cosmic censorship conjecture. However, for regular RBHs it has been suggested that it could be possible to destroy the event horizon \cite{Li&Bambi}.

A warning is relevant here: A RBH solution should be stable in the region outside the event horizon and also inside. In the previous considerations (and figures) we have not taken into account the effects that instabilities could have in the global structure of the spacetime. Precisely, a non-trivial problem for RBH is the stability of their inner horizon. The first works in instability of inner horizons come from the study of the classical charged Reissner-Nordstr\"{o}m black hole which suffers the so called \textit{mass-inflation instability} \cite{P&I}. Studies of the instability of Kerr's  horizons were developed in \cite{B&W}\cite{P&I2}. The consideration of regular black holes coming from different approaches to quantum gravity does not seem to alleviate the problem since, first, even in the non-rotating case they seem to require the existence of an (usually unstable) inner horizon and, second, even the backscattered flux of Hawking radiation coming from the black hole itself could be enough to destabilize its inner horizon \cite{TorresIns}. Other studies on the stability of regular black holes can be found in \cite{Cetal1}\cite{Cetal2}\cite{Cetal3}. Even a fine-tuned stable regular RBH can be found in \cite{FLMV}.

\section{Causality}\label{secCaus}

In general, it seems reasonable to ask a time orientable spacetime to be absent of closed causal curves. The existence of such curves would seem to lead to logical paradoxes: One could travel following these curves and arrive back before one's departure, so that one could prevent oneself from setting out in the first place.
A spacetime absent of closed causal curves is said to be \textit{causal} \cite{H&E}. If, in addition, no closed causal curve appears even under any small perturbation of the metric the spacetime is called \textit{stably causal}.
It is well-known that the usual analytical extension of the Kerr metric is non-causal. Since Kerr metric is a particular case of the metric (\ref{gIKerr}), it is natural to ask whether the maximal extensions of regular RBH should also be non-causal.

Along the lines in \cite{Maeda}, in order to examine this issue we will use proposition 6.4.9 in \cite{H&E} that states that when a time function $f$ exists in the spacetime such that its normal $\mathbf{n}\equiv\nabla_\mu f \, dx^\mu$ is timelike, then the spacetime is stably causal. ($f$ can be thought as \emph{the} time in the sense that it increases along every future-directed causal curve).
Let us choose the time coordinate $\tilde{t}$ in Kerr-Schild coordinates as our time function $f$. The timelike character of $\mathbf{n}$ can be checked as follows:
\begin{equation}\label{n2}
\mathbf{n}^2=g^{\mu\nu} \nabla_\mu \tilde{t} \nabla_\nu \tilde{t}=g^{\tilde{t}\tilde{t}}=-1-\frac{2 \mathcal M(r) r^3}{r^4+a^2 z^2}.
\end{equation}
Since we would like this to be negative, it trivially follows
\begin{theorem} \cite{Maeda}
If $r \mathcal M(r)\geq 0$ for all $r$, then the model of RBH with metric (\ref{GGg}) [or (\ref{gIKerr})] will be stably causal.
\end{theorem}
Note that for a regular RBH (unextended through $r=0$), a non-negative mass function suffices to guarantee a stably causal spacetime.

\section{Thermodynamics}

The consideration of the thermodynamics of black holes started in the 1970's with a series of articles with fundamental contributions by Bekenstein and Hawking \cite{BCH}\cite{beke}. In 1975, Hawking proved that quantum mechanical effects cause Schwarzschild black holes to create and emit particles as if they were black bodies with a temperature proportional to their surface gravity. Since then, the thermodynamics of many different black holes coming from General Relativity and from alternative theories has been analyzed.
Here we would like to treat the thermodynamics of general regular RBH described by the line element (\ref{gIKerr}) at an introductory level.

In a RBH there is a family of \textit{stationary observers}, i.e., those observers moving with constant angular velocity at fixed $r$ and $\theta$ without perceiving any time variation of the gravitational field. It is easy to check that their angular velocity is
\[
\Omega=\frac{d\phi}{dt}=-\frac{g_{t\phi}}{g_{\phi\phi}}=\frac{a (a^2+r^2-\Delta)}{(r^2+a^2)^2-\Delta a^2 \sin^2\theta}.
\]
The four velocity of these observers is proportional to the killing vector
\[
\vec \xi\equiv \vec t+\Omega \vec\phi,
\]
constructed with the killing vectors $\vec t=\partial_t$ and $\vec\phi=\partial_\phi$ and the constant ($r$ and $\theta$ fixed for the stationary observers) angular velocity $\Omega$.
In the event horizon ($r_+$) the angular velocity is just
\[
\Omega(r_+)=\Omega_+=\frac{a}{r_+^2+a^2}
\]
and it can be checked that the kiliing vector on the horizon $\vec \xi\rfloor_{r_+}$ is light-like  ($\vec \xi\cdot\vec \xi\rfloor_{r_+}=0$). In this way, this killing is tangent to the null geodesics generators of the event horizon.

The surface gravity $\kappa$ in the horizon can be found using \cite{Wald}
\[
\kappa^2=-\frac{1}{2} \nabla_\mu \xi_\nu\, \nabla^\mu \xi^\nu.
\]
In our regular RBH it is just
\[
\kappa=\frac{\Delta'(r_+)}{2 (r_+^2+a^2)},
\]
where the prime in $\Delta'$ stands for derivative with respect to $r$. Note that the surface gravity is constant in the event horizon, what links it to the temperature of the RBH. In fact, it is usually assumed that the temperature is just $T_+=\kappa/2\pi$ \cite{Wald} and, thus,
\[
T_+=\frac{\Delta'(r_+)}{4 \pi (r_+^2+a^2)}.
\]
In this way, whenever $\Delta'(r_+)\neq 0$ the black hole will have a non-zero temperature and will emit Hawking radiation.
Note also that a differentiable $\Delta$ requires, in the extremal case, that $\Delta'(r_\pm)= 0$. In this way, in case it exists, the temperature of an extremal RBH would be zero and it would not emit Hawking radiation.

In order to check the correctness of the result, one can compute the temperature of the particular case $\mathcal M(r)=m=$constant (i.e., the well-known Kerr's RBH) obtaining the expected result
\[
T^{Kerr}_+=\frac{r_+^2-a^2}{4\pi r_+ (r_+^2+a^2)}.
\]

Further discussions of the thermodynamic properties and Hawking radiation for particular regular RBHs can be found in \cite{A&G2}\cite{HSK}\cite{R&T}\cite{take}.

\section{Obtaining Regular Rotating Black Hole Models}\label{secObt}

Different articles dealing with regular rotating black holes propose different forms for the function $\mathcal M(r)$. Its exact expression depends on the procedure used to obtain the RBH.
In many cases the authors just propose heuristic forms for $\mathcal M$.
The idea behind this heuristic approach is to try to ascertain the main characteristics that a regular RBH should have. Thus, for instance, the possible differences between the event horizons in the proposed models and the event horizon in Kerr's solution can be analyzed and maybe observationally tested (see section \ref{secPheno}).
Of course, for a regular RBH the specific mass function $\mathcal M$ is chosen to avoid the existence of singularities. It is usually also demanded that the spacetime should be asymptotically flat. Other goals may include the (approximated) fulfillment of energy conditions beyond the event horizon, the stability of the model \cite{FLMV} or a good causal behaviour of the model. (See, for instance, \cite{A-A}\cite{B&M}\cite{LGS}\cite{Maeda}\cite{MFL}\cite{eye}). Even $d$-dimensional ($d>4$) regular RBH have been studied heuristically. (See, for instance, \cite{A&G}\cite{Amir}).

In other cases a physical approach provides a specific $\mathcal M(r)$.
Let us just mention a few of them. Some authors, inspired by the work of Bardeen \cite{Bardeen}, have taken the path of nonlinear electrodynamics, which provides the necessary modifications in the energy-momentum tensor in order to avoid singularities in the RBH \cite{D&G}\cite{Ghosh}\cite{Tosh}. Yet, another way of addressing the problem of singularities is to take into account that quantum gravity effects should play an important role in the core of black holes, so that it would seem convenient to directly derive the black hole behaviour from an approach to quantum gravity. In this way, regular RBHs deduced in the Quantum Einstein Gravity approach can be found in \cite{R&T}\cite{TorresExt}, in the framework of Conformal Gravity in \cite{BMR}, in the framework of Shape Dynamics in \cite{G&H}, inspired by Supergravity in \cite{Buri}, by Loop Quantum Gravity in \cite{C&M} and by non-commutative gravity in \cite{S&S}.

In the case of non-heuristic models, theoretically one \emph{obtains} a specific expression for the mass function and then one has to check for the avoidance of singularities and for the rest of desirable properties cited above.
The study of their event horizon is particularly important here since it may be observationally tested in the future (see section \ref{secPheno}), what could, for instance, help selecting among the different candidates to a Quantum Gravity Theory.

\subsection{Generalized Newman-Janis Algorithms}

In 1965, Newman and Janis \cite{N&J} discovered that it was possible to obtain Kerr's solution by applying an algorithm to a spherically symmetric and static \emph{seed} metric: Schwarzschild's solution. A great step towards understanding the algorithm, its possibilities, generalizations and limitations was carried out in \cite{Drake&Szek}. The generalized algorithm allows to take any static spherically symmetric seed metric and obtain a rotating axially symmetric offspring from it\footnote{The reader should be aware that the \textit{offspring} has different geometrical properties and also different physical properties. For example, the seed metric can be a perfect fluid, but the offspring will never be another perfect fluid \cite{Drake&Szek}.}. The application to regular RBH followed \cite{B&M}: One starts with a regular static and spherically symmetric black hole from a specific framework. Then, one applies the generalized N-J algorithm to try to construct a regular RBH.

The generalized Newman-Janis algorithm is a five-step procedure \cite{Drake&Szek}:
\begin{enumerate}
\item Take a static spherically symmetric line element and write it in advanced null coordinates.
\item Express the contravariant form of the metric in terms of a null tetrad $Z^\mu_a$.
\item Extend the coordinates $x^\rho$ to a new set of complex coordinates
\[
x^\rho \rightarrow \tilde{x}^\rho=x^\rho+i y^\rho(x^\sigma)
\]
and let the null tetrad vectors $Z^\mu_a$ undergo a transformation
\[
Z^\mu_a \rightarrow \tilde{Z}^\mu_a(\tilde{x}^\rho, \bar{\tilde{x}}^\rho).
\]
Require that the transformation recovers the old tetrad and metric when $\tilde{x}^\rho=\bar{\tilde{x}}^\rho$.
\item Obtain a new metric by making a complex coordinate transformation
\[
\tilde{x}^\rho=x^\rho+i \gamma^\rho(x^\sigma)
\]
\item Apply a coordinate transformation $u=t+\mathcal F(r)$, $\phi=\varphi+\mathcal H(r)$ to transform the metric to Boyer-Lindquist-type coordinates.
\end{enumerate}

While the seed spacetime can be a general spherically symmetric static spacetime, we will restrict ourselves here to a specific family of seed spacetimes that will allow us to connect with our family of regular RBH (\ref{gIKerr}).
Lets say that one has found a line element for a static regular spherically symmetric black hole (in a determined framework or just in a heuristic manner). Assume that the found line element can be written in coordinates $\{t,r,\theta,\varphi\}$ as a member of the family of static regular spherically symmetric black holes with metric:
\begin{equation}\label{seed}
ds^2=-f(r) dt^2+f^{-1}(r) dr^2+r^2 d\Omega^2,
\end{equation}
where $d\Omega^2=d\theta^2+\sin^2\theta d\varphi^2$.
The function $f(r)$ can be rewritten as
\[
f(r)=1- 2 \frac{ M (r)}{r},
\]
by using the \textit{mass function} $M (r)$ defined for general spherically symmetric spacetimes \cite{M&S}.

Let us now see that \emph{the generalized N-J algorithm provides us with a means of obtaining the corresponding rotating black hole line element (\ref{gIKerr}) from the static spherically symmetric seed metric (\ref{seed})}.

The specific five steps for this case would be:
\begin{enumerate}
\item
The coordinate change $du=dt+dr/f(r)$ allows us to rewrite the metric in advanced null coordinates\footnote{Note that in the literature on the NJ algorithm there is some confusion between the advanced and the retarded ($dw=dt-dr/f(r)$) null coordinates. The first is suitable for describing black holes, the second for white holes.}
\[
ds^2=-f(r) du^2+2 du dr + r^2 d\Omega^2.
\]
\item
The null tetrad $Z^\mu_a=(l^\mu,n^\mu,m^\mu,\bar m^\mu)$ satisfying
$l_\mu n^\mu=-m_\mu\bar m^\mu=-1$ and $l_\mu m^\mu=n_\mu m^\mu=0$
can be chosen as
\[
l^\mu=-\delta^\mu_r,\hspace{1 cm} n^\mu=\delta^\mu_u+\frac{f(r)}{2}\, \delta^\mu_r,\hspace{1 cm}
m^\mu=\frac{1}{\sqrt{2} r} \left(\delta^\mu_\theta+\frac{i}{\sin\theta} \delta^\mu_\varphi\right)
\]
so that $g^{\mu\nu}=-l^\mu n^\nu-l^\nu n^\mu+m^\mu \bar m^\nu+m^\nu \bar m^\mu$. (Note that both $\vec l$ and $\vec n$ are future directed).
\item We perform the coordinate change
\[
r'=r-i\, a \cos\theta, \hspace{1 cm}  u'=u-i\, a \cos\theta.
\]
and demand $r'$ and $u'$ to be real.
In this way the null tetrad transforms into ($Z'^\mu_a=Z^\nu_a \partial x^{\mu'}/\partial x^\nu$)
\[
l'^\mu=-\delta^\mu_r, \hspace{.5cm} n'^\mu=\delta^\mu_u+\frac{\bar f(r')}{2}\, \delta^\mu_r,\hspace{.5 cm}
m'^\mu=\frac{1}{\sqrt{2} r'} \left(\delta^\mu_\theta+\frac{i}{\sin\theta} \delta^\mu_\varphi+i\, a \sin\theta (\delta^\mu_u+\delta^\mu_r)\right)
\]
The function $\bar f$ comes from the complexification of $f$ and, for the moment, we only know that it must be real and that it must reproduce Kerr solution if the complexified mass function is just a constant. This is possible if, as usual \cite{B&M}\cite{Drake&Szek}\cite{N&J}, one uses the complexification
\[
\frac{1}{r}\rightarrow \frac{1}{2} \left(\frac{1}{r'}+\frac{1}{\bar r'} \right),
\]
that provides us with
\begin{equation}
\bar f=1- \frac{2 \bar{ M} (r,\theta) r}{\Sigma},\label{barf}
\end{equation}
where there is still some freedom in choosing the function $\bar{ M}(r, \theta)$.

\item
The new non-zero metric coefficients can be computed to be
\begin{eqnarray}\label{metcoef}
g_{uu}&=&-\bar f(r,\theta), \hspace{.5cm} g_{ur}=+1, \hspace{.5cm} g_{u\varphi}=-a \sin^2\theta [1-\bar f (r,\theta)]\\
g_{r\varphi}&=&-a \sin^2 \theta, \hspace{.5cm} g_{\theta\theta}=\Sigma, \hspace{.5cm} g_{\varphi\varphi}=\sin^2\theta [\Sigma +a^2 \sin^2\theta (2-\bar f)]\nonumber
\end{eqnarray}
\item In order to get the metric in Boyer-Lindquist type coordinates $\{u,r,\theta, \phi \}$ we perform the coordinate change
$u=t+\int F(r) dr$, $\varphi=\phi+\int H(r) dr$,
where
\begin{equation}\label{transBL}
F(r)=\frac{r^2+a^2}{\bar f(r,\theta) \Sigma +a^2\sin^2 \theta}\ \  \mbox{and}\ \
H(r)=\frac{a}{\bar f(r,\theta) \Sigma+a^2\sin^2 \theta}.
\end{equation}
Thus, $\bar f$ should be chosen in such a way that $F$ and $H$ must be functions of $r$ alone.
Note that (\ref{transBL}) implies
\[
\Sigma \bar f(r,\theta)+a^2 \sin^2\theta=D(r),
\]
and substituting $\bar f $ using (\ref{barf}) one immediately sees that this step requires $\bar{M}=\bar{M}(r)$, i.e., $\bar{ M}$ cannot depend on $\theta$. Thus, we arrive at the natural choice $\bar{ M}(r)= M(r)$.
In effect, in this case $F$ and $H$ are really functions of $r$ alone since
\[
F(r)=\frac{r^2+a^2}{r^2+a^2-2 \mathcal M(r) r} \ \  \mbox{and}\ \
H(r)=\frac{a}{r^2+a^2-2 \mathcal M(r) r}
\]
(see (\ref{uphi})). Therefore, in this way it is possible to write the solution in Boyern-Lindquist type coordinates as

\begin{equation}\label{metRBH}
ds^2=-\frac{\Delta}{\Sigma} (dt-a \sin^2\theta d\phi)^2+\frac{\Sigma}{\Delta} dr^2+
\Sigma d\theta^2+\frac{\sin^2\theta}{\Sigma} (a dt-(r^2+a^2) d\phi)^2.
\end{equation}
where the mass fuction $M(r)$ (appearing in $\Delta$) should just be relabelled as $\mathcal M(r)$ in order to be exactly (\ref{gIKerr}).
\end{enumerate}

Since we started with a \emph{regular} spherically symmetric seed spacetime, let us assume that its mass function satisfies $M(r)= O(r^n)$ with $n\geq 3$ around $r=0$. Thus, its offspring also has a mass function $\mathcal M (r)=M(r)= O(r^n)$ and, therefore, is also devoid of scalar polynomial curvature singularities (according to theorem \ref{teorema}).

The moral of the procedure is that, if one wants a regular RBH in B-L-like coordinates, then one can simply take the mass function obtained in a spherically symmetric framework with metric (\ref{seed}) and use it as the mass function in the metric (\ref{gIKerr}).

Note that, if one considers step 5 (obtaining the line element in B-L-like coordinates) as non-compulsory, then one could analyze the complexification for two different cases \cite{B&M}:
\begin{enumerate}
\item \textit{Type I} in which we impose $\mathcal M=\mathcal M(r)$. This is the case that we have just considered and the usual approach in the literature.
\item \textit{Type II} in which we allow $\mathcal M=\mathcal M(r,\theta)$. The new rotating metric can be written in Kerr form (with a null coordinate in the style of Eddington-Finkelstein coordinates), but the N-J algorithm cannot be completed since it is not possible to write the rotating metric in the final Boyer-Lindquist form. Specific models of this type have been proposed and explored in \cite{B&M}\cite{E&H}\cite{EH1}.
\end{enumerate}

\section{Phenomenology}\label{secPheno}

Recent developments have greatly enhanced our ability to probe theoretical predictions concerning black holes. These include, in the one hand, the direct observation of gravitational waves emanating from astrophysical sources by the LIGO-VIRGO-KAGRA collaborations \cite{ligo} and, in the other hand, the images of black holes taken by the Event Horizon Telescope (EHT) \cite{Aki}\cite{Aki2}. Moreover, a considerable enhancement is expected in the near future thanks to the LISA project \cite{lisa} and the new planned ground-based observatories \cite{observ}. In this way, the physics in strong gravitational fields near black holes is becoming an important topic not only in theoretical physics but also in astrophysical phenomenology. There is now a need for compiling the maximum amount of theoretical results about realistic rotating black holes. It is hoped that the phenomenological evidence will help us to choose among the different proposals for rotating black hole models and, as a consequence, among the alternative approaches to gravitational theories.

\subsection{Shadows}

A defining characteristic of a black hole is the event horizon. To a distant observer, the event horizon casts a relatively large "shadow" with, according to General Relativity,  an apparent diameter of $\sim 10$ gravitational radii that is due to the bending of light by the black hole.
Of course, the specific theoretical characteristics of this shadow depend on the alternative gravitational theory chosen and the properties of the modeled rotating black hole. Currently, there are numerous studies about RBH shadows in different frameworks for alternative theories. Just to mention a few: heuristic approaches can be found in \cite{AAAG}\cite{ASG}\cite{Am&G}\cite{E&H}\cite{LGPV}\cite{L&B}\cite{S&S0}, results from different approaches to Quantum Gravity can be found in \cite{BCY}\cite{HGE}.

Assuming a metric of the G\"{u}rses-G\"{u}rsey type (\ref{gIKerr}), a general formula for the contour of the shadow can be easilly obtained \cite{Tsuka}:  The photons that would define the shadow of the regular rotating black hole are described by an action $S=S(x^\alpha)$. The momentum of the photons is
\[
p_{\mu}\equiv \frac{\partial S}{\partial x^\mu}
\]
and satisfy
\begin{equation}\label{photon}
g^{\alpha\beta} p_\alpha p_\beta=0.
\end{equation}
The stationary and axisymmetric of the spacetime described by the metric (\ref{gIKerr}) imply two conserved quantities in the trajectory of the photon: the energy $E\equiv -p_t$ and the angular momentum $L\equiv p_\phi$. If there is a separable solution for $S$, by using the definition of the momentum, we could rewrite it as
\[
S=-E t+L \phi+ S_r+ S_\theta,
\]
where we have introduced the new functions $S_r=S_r(r)$ and $S_\theta=S_\theta(\theta)$. In this way, (\ref{photon}) can now be written as
\begin{equation}\label{SrSth}
-\Delta \left(\frac{d S_r}{dr} \right)^2+ \frac{[(r^2+a^2) E-a L]^2}{\Delta}=\left(\frac{d S_\theta}{d\theta} \right)^2+ \frac{(L-a E \sin^2\theta)^2}{\sin^2\theta}.
\end{equation}
In this equation, the left-hand side depends only on $r$, while the right-hand side depends only on $\theta$. In this way, they define a constant which we will denote by
\begin{equation}\label{KSth}
K=\left(\frac{d S_\theta}{d\theta} \right)^2+ \frac{(L-a E \sin^2\theta)^2}{\sin^2\theta}.
\end{equation}
From $dx^\mu/d\lambda=p^\mu=g^{\mu\nu} p_\nu$ and using (\ref{SrSth}) and (\ref{KSth}) one gets
\begin{equation}\label{evolR}
\Sigma \frac{dr}{d\lambda}=\pm \sqrt{R(r)},
\end{equation}
where $R(r)\equiv P(r)^2-\Delta [(L-a E)^2+\mathcal Q]$, $P(r)\equiv E (r^2+a^2)-a L$ and $\mathcal Q\equiv K-(L-a E)^2$ is the Carter constant.

Equation (\ref{evolR}) implies that there would be  unstable circular orbits at a certain $r=r_0$ whenever $R(r_0)=R'(r_0)=0$ and $R''(r_0)>0$.
To exploit this, note that the definition of $R$ can also be rewritten as
\[
R/E^2=r^4+(a^2-\xi^2-\eta) r^2 + 2 \mathcal M (r) [(\xi-a)^2+\eta] r-a^2 \eta,
\]
where $\eta\equiv \mathcal Q/E^2$ and $\xi\equiv L/E$. The derivative of this expression with respect to $r$ provides
\[
R'/E^2=4 r^3+2 (a^2-\xi^2-\eta) r+ 2 \mathcal M (r) [(\xi-a)^2+\eta] f(r),
\]
where
\[
f(r)\equiv 1+\frac{r \mathcal M' }{\mathcal M}.
\]
Using the conditions for the orbit one gets the quadratic equation with respect to $\xi$:
\begin{align*}
a^2 &(r_0-f_0 \mathcal M_0)\xi^2-2 a \mathcal M_0 [(2-f_0) r_0^2-f_0 a^2] \xi-r_0^5+\\
 &+(4-f_0) \mathcal M_0 r_0^4-2 a^2 r_0^3+2 a^2 \mathcal M_0 (2-f_0) r_0^2 -a^4 r_0-a^4 \mathcal M_0 f_0=0,
\end{align*}
where $\mathcal M_0\equiv\mathcal M(r_0)$ and $f_0\equiv f(r_0)$.
In order to describe the black hole shadow, we must choose the solution
\[
\xi_-\equiv\frac{4 \mathcal M_0 r_0^2-(r_0+f_0 \mathcal M_0)(r_0^2+a^2)}{a (r_0-f_0 \mathcal M_0)}.
\]
that implies
\[
\eta=\eta_-\equiv \frac{r_0^3 [4 (2-f_0) a^2 \mathcal M_0-r_0 [r_0-(4-f_0)\mathcal M_0]^2]}{a^2 (r_0-f_0 \mathcal M_0)^2}.
\]
We consider an observer at a large distance from the RBH in the asymptotically flat spacetime thats observes the RBH with an inclination $\theta_i$. The contour of the shadow of the black hole can be expressed by celestial coordinates $\alpha$ and $\beta$ \cite{Tsuka} as
\[
\alpha=\frac{-\xi_-}{\sin\theta_i}\hspace{.5 cm}; \hspace{.5 cm} \beta= \pm \sqrt{\eta_-+(a-\xi_-)^2-\left(a \sin \theta_i-\frac{\xi_-}{\sin\theta_i} \right) ^2} .
\]

\begin{figure}[htp]
\includegraphics[scale=.7]{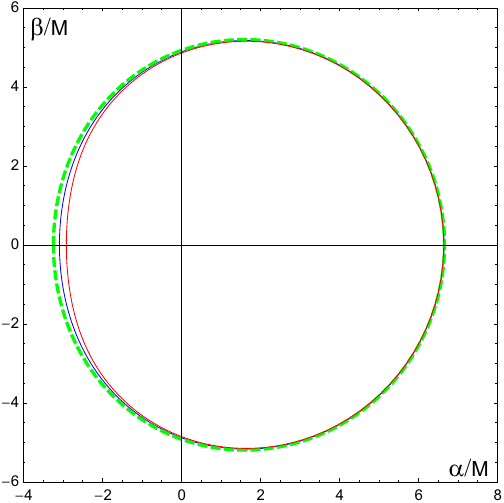}
\caption{\label{shadows} Shadows of RBHs for three different cases. We have fixed the rotation parameter such that $a/M=0.8$ and the inclination angle $\theta_i=\pi/2$ (BH seen from the equatorial plane). The dashed green shadow corresponds to Kerr's case. The blue shadow corresponds to Hayward's RBH with $g/M=0.5$. The red shadow corresponds to Hayward's RBH with $g/M=0.6$. Note that for smaller $g/M$ the deviations from Kerr's shadow are smaller. In this way, for Hayward's RBH and $g$ of the order of Planck's mass the deviation of the shadow from the classical case is negligible.}
\end{figure}

In this way, we have finally arrived at the expressions that link the parameters describing the RBH with the observer's celestial coordinates. Using these expressions, in figure \ref{shadows} we have plot the shadows of regular RBHs compared to the shadow of Kerr's RBHs. In the example we have chosen, for the sake of simplicity, Hayward's regular RBH \cite{Hay2006} with a mass function
\[
\mathcal M(r)=\frac{r^3}{r^3+g^3} M,
\]
where $M$ is a constant (the mass of the RBH as measured by observers at infinity) and $g$ is a positive constant with dimensions of mass that quantifies the deviation from Kerr's solution.

The current observations of black hole shadows by the Event Horizon Telescope are so far consistent with the shadow predicted for Kerr's RBHs. However, this classical solution of Einstein's equations cannot provide a complete understanding of black holes since, for instance, it implies the existence of an inner singularity. An analysis of the regular RBH studied so far shows that their shadows are usually also compatible with the observed shadows. In fact, the shadows of the regular RBH are indistinguishable from Kerr black holes shadows within the current observational uncertainties \cite{KKG}\cite{LGPV}\cite{L&B}. Future mm/sub-mm VLBI facilities will be able to greatly increase the current observational resolution. Even so, it will be challenging to test these metrics in the near future.
A primary reason for this is (as explained in section \ref{Horizons}) that only slightly more compact event horizons and smaller shadows are usually expected. In fact, if the deviation from General Relativity comes from Quantum Gravity effects and it is Planck's scale which provides us with the scale in which to expect the departure, then it would be practically impossible to observe these effects in the shadows of a massive rotating black hole. A better prospect for future observations would be expected if, on the contrary, the scale could be much bigger, as pointed out by some authors \cite{Dvali}\cite{Mathur}, or the resolution of singularities were not related to Quantum Gravity effects.

\section{Summary}\label{conclu}


Assuming that a manifold endowed with its corresponding metric is a fairly good approximation for describing a regular RBH, most of the models in the literature are of the Gürses-Gürsey type, whose general properties have been discussed in this text. We have seen that the regularity condition for these models translates into a condition for their mass function. We showed that the requirement of regularity leads to the violation of the energy conditions. Remarkably, regular RBH do not seem to require an extension through their disk. In this way, causality problems could be avoided simply if their mass function could remain non-negative. With regard to the choice of their mass function, in the literature it has been either chosen heuristically or derived from some gravitational theory. In either case, the generalized Newman-Janis algorithm provides us with an alibi to just use the mass function from regular spherically symmetric static black hole models.

The recent observational developments (LIGO-VIRGO-KAGRA collaborations, the Event Horizon Telescope or, in the near future, the LISA project) opened the possibility to probe our theoretical predictions on rotating black holes. With our current observational resolution, so far the observations are consistent both with General Relativistic black holes and with alternative regular RBH models. Nevertheless,
there is hope that a future increase in observational resolution could lead to discern among the different theoretical predictions.

There are some open questions concerning regular RBHs. For example, the analysis of the possible existence of parallelly propagated curvature singularities, the treatment and resolution of their inner horizon instabilities, their future evolution due to the emission of Hawking radiation (including the problem of the possible formation of remnants) and a deeper generalization of the models beyond the Gürses-Gürsey type.




\begin{thebibliography}{99}

\bibitem{AAAG}
Abdujabbarov A, Amir M, Ahmedov B and Ghosh SG 2016 {\it Phys. Rev. D} {\bf 93} 104004 (arXiv:1604.03809 [gr-qc])

\bibitem{ASG}
Ahmed F, Singh DV and Ghosh SG 2022 {\it Gen. Rel. and Grav.} {\bf 54} 21 	(arXiv:2002.12031 [gr-qc])

\bibitem{Aki}
Akiyama K \textit{et al.} [Event Horizon Telescope] 2019 {\it Astrophys. J. Lett} {\bf 875} L1, L2, L3, L4, L5, L6

\bibitem{Aki2}
Akiyama K \textit{et al.} [Event Horizon Telescope] 2022 {\it Astrophys. J. Lett} {\bf 930} L12, L13, L14, L15, L16, L17

\bibitem{A&G}
Ali MS and Ghosh SG 2018 {\it Phys. Rev. D} {\bf 98} 084025

\bibitem{A&G2}
Ali MS and Ghosh SG 2019 {\it Phys. Rev. D} {\bf 99} 024015

\bibitem{Am&G}
Amir M and Ghosh SG 2016 {\it Phys. Rev. D} {\bf 94} 024054 (arXiv:1603.06382 [gr-qc])

\bibitem{Amir}
Amir M, Ali MS and Maharaj SD 2020 {\it Class. Quantum Grav.} {\bf 37} 145014

\bibitem{A&B2005}
Ashtekar A and Bojowald M 2005 {\it Class. Quantum Grav.} {\bf 22} 3349

\bibitem{A-BI}
Ayon-Beato E and Garcia A 2000 {\it Phys. Lett. B} {\bf 493} 149

\bibitem{A-BII}
Ayon-Beato E and Garcia A 2005 {\it Gen. Rel. and Grav.} {\bf 37} 635

\bibitem{A-A}
Azreg-A\"{\i}nou M 2014 {\it Phys. Rev. D} {\bf 90} 064041 (arXiv:1405.2569 [gr-qc])

\bibitem{B&V}
Balart L and Vagenas E C 2014 {\it Phys. Rev. D} {\bf 90} 124045

\bibitem{B&M}
Bambi C and Modesto L 2013 {\it Phys. Lett. B} {\bf 721} 329 (arXiv:1302.6075 [gr-qc])

\bibitem{BMR}
Bambi C, Modesto L and Rachwal L 2017 {\it JCAP} {\bf 05} 003 (arXiv:1611.00865 [gr-qc])

\bibitem{lisa}
Barausse E \textit{et al.} 2020 {\it Gen. Rel. and Grav.} {\bf 52} 81

\bibitem{Bardeen}
Bardeen J M 1968 {\it Conference Proceedings of GR5} p.174

\bibitem{BCH}
Bardeen JM, Carter B and Hawking SW 1973 {\it Commun. Math. Phys.} {\bf 31} 161

\bibitem{beke}
Bekenstein JD 1972 {\it Lett. Nuovo Cimento} {\bf 4} 737

\bibitem{B&W}
Bernard FW 1989 {\it J. Math. Phys.} {\bf 30} 1301

\bibitem{B&R}
Bonanno A and Reuter M 2000 {\it Phys. Rev. D} {\bf 62} 043008 (arXiv:0002196 [hep-th])

\bibitem{B&L}
Boyer R H and Lindquist R W 1967 {\it J. Math. Phys.} {\bf 8} 265

\bibitem{BCY}
Brahma S, Chen C-Y and Yeom D-H 2021 {\it Phys. Rev. Lett.} {\bf 126} 181301 (arXiv:2012.08785 [gr-qc])

\bibitem{Buri}
Burinskii A 2002 {\it Czech.J.Phys. } {\bf 52} C471

\bibitem{C&M}
Caravelli F and Modesto L 2010 {\it Class.Quant.Grav.} {\bf 27} 245022 (arXiv:1006.0232 [gr-qc])

\bibitem{Cetal1}
Carballo-Rubio R, Di Filippo F, Liberati S, Pacilio C and Visser M 2018 {\it JHEP} {\bf 07} 023 (arXiv:1805.02675 [gr-qc])

\bibitem{Cetal2}
Carballo-Rubio R, Di Filippo F, Liberati S, Pacilio C and Visser M 2021 arXiv: 2101.05006 [gr-qc]

\bibitem{Cetal3}
Carballo-Rubio R, Di Filippo F, Liberati S, Pacilio C and Visser M 2022 arXiv: 2205.13556 [gr-qc]

\bibitem{Carter1968a}
Carter B 1968 {\it Phys. Rev.} {\bf 174} 1559

\bibitem{LGS}
De Lorenzo T, Giusti A and Speziale S 2016 {\it Gen. Rel. and Grav.} {\bf 48} 31. Corrigendum at {\it Gen. Rel. and Grav.} {\bf 48} 111

\bibitem{Dono}
Donoghue JF 1994 {\it Phys. Rev. Lett.} {\bf 72} 2996; 1994 {\it Phys. Rev. D} {\bf 50} 3874

\bibitem{Drake&Szek}
Drake S P and Szekeres P 2000 {\it Gen. Rel. and Grav.} {\bf 32} 445

\bibitem{Dvali}
Dvali G and Gomez C 2011 	arXiv:1112.3359 [hep-th]

\bibitem{Dymni}
Dymnikova I 2004 {\it Class. Quantum Grav.} {\bf 21} 4417 (arXiv:0407072 [gr-qc])

\bibitem{D&G}
Dymnikova I and Galaktionov E 2015 {\it Class. Quant. Grav.} {\bf 32} 165015 (arXiv:1510.01353 [gr-qc])

\bibitem{E&H}
Eichhorn A and Held A 2021 {\it JCAP} {\bf 05} 073 	(arXiv:2103.13163 [gr-qc])

\bibitem{EH1}
Eichhorn A and Held A 2021 {\it Eur. Phys. J C} {\bf 81} 933 (arXiv:2103.07473 [gr-qc])

\bibitem{FLMV}
Franzin E, Liberati S, Mazza J and Vellucci V 2022 arXiv:2207.08864 [gr-qc]

\bibitem{Frolov2014}
Frolov V P 2014 {\it JHEP} {\bf 5} 049 (arXiv:1402.5446 [hep-th])

\bibitem{G&P2014}
Gambini R and Pullin J 2015 {\it Proceedings of Science} {\bf 14} 038 (arXiv:1408.3050 [gr-qc])

\bibitem{GC&M}
Garc\'{\i}a-Compe\'{a}n H and Manko VS 2015 {\it Prog. Theor. Exp. Phys.} {\bf 2015} 043E02 (arXiv:1205.5848 [gr-qc])

\bibitem{Ghosh}
Ghosh S G 2015 {\it Eur. Phys. J. C} {\bf 75} 532 (arXiv:1408.5668 [gr-qc])

\bibitem{G&V}
Gibbons GW and Volkov MS 2017 {\it Phys. Rev. D} {\bf 96} 024053 (arxiv:1705.07787 [hep-th])

\bibitem{G&H}
Gomes H and Herczeg G 2014 {\it Class.Quant.Grav.} {\bf 31} 175014 (arXiv:1310.6095 [gr-qc])

\bibitem{Griff}
Griffths J B and Podolsk\'{y} J 2009 {\it Exact space-times in Einstein's General Relativity} Cambridge: Cambridge University Press.

\bibitem{GG}
G\"{u}rses M and G\"{u}rsey F 1975 {\it Phys. Rev. D} {\bf 11} 967

\bibitem{H&R2014}
Haggard H M and Rovelli C 2015 {\it Phys. Rev. D} {\bf 92} 104020 (arXiv:1407.0989 [gr-qc])

\bibitem{Hamity}
Hamity V 1976 {\it Phys. Lett. A} {\bf 56} 77

\bibitem{H&E}
Hawking SW and Ellis GFR 1973 {\it The large scale structure of space-time} Cambridge University Press, Cambridge

\bibitem{Hay2006}
Hayward S A 2006 {\it Phys. Rev. Lett.} {\bf 96} 031103

\bibitem{HGE}
Held A, Gold R and Eichhorn A 2019 {\it JCAP} {\bf 1906} 029 (arXiv:1904.07133 [gr-qc])

\bibitem{HSK}
Hendi SH, Sajadi SN and Khademi M 2021 {\it Phys. Rev. D} {\bf 103} 064016 (arXiv:2006.11575 [gr-qc])

\bibitem{IsraelThin}
Israel W 1966 {\it Nuovo Cimento B} {\bf 44} 1 ; 1967 {\it Nuovo Cimento B} {\bf 48} 463.

\bibitem{Israel}
Israel W 1970 {\it Phys. Rev. D} {\bf 2} 641

\bibitem{observ}
Kalogera V \textit{et al.} arXiv:2111.06990 [gr-qc]

\bibitem{KKG}
Kumar R, Kumar A and Ghosh SG 2020 {\it Astrophys. J} {\bf 896}:89

\bibitem{LGPV}
Lamy F, Gourgoulhon E Paumard T sand Vincent FH  2018 {\it Class. Quantum Grav.} {\bf 35} 115009 (arXiv:1802.01635 [gr-qc])

\bibitem{Li&Bambi}
Li Z and Bambi C. 2013 {\it Phys.Rev.D} {\bf 87} 124022 (arXiv:1304.6592 [gr-qc])

\bibitem{L&B}
Li Z and Bambi C 2014 {\it JCAP} {\bf 1401} 041 (arXiv:1309.1606 [gr-qc])

\bibitem{Maeda}
Maeda H 2021 (arXiv:2107.04791 [gr-qc])

\bibitem{Mathur}
Mathur SD 2005 {\it Fortsch.Phys.} {\bf 53} 793	(arXiv:0502050 [hep-th])

\bibitem{MFL}
Mazza J, Franzin E and Liberati S 2021 {\it JCAP} {\bf 04} 082

\bibitem{M&S}
Misner CW and Sharp DH 1964 {\it Phys. Rev.} {\bf 136} B571

\bibitem{gravit}
Misner CW, Thorne KS and Wheeler JA 1970 {\it Gravitation},
W.H. Freeman and Company, New York

\bibitem{N&J}
Newman, E T and Janis A I 1965 {\it J. Math. Phys.} {\bf 6} 915

\bibitem{P&C}
Pani P and Cardoso V 2009 {\it Phys. Rev. D} {\bf 79} 084031 (arXiv:0902.1569 [gr-qc])

\bibitem{P&I2}
Poisson E  and Israel W 1989 {\it Phys. Rev. Lett.} {\bf 63} 1663

\bibitem{P&I}
Poisson E and Israel W 1990 {\it Phys. Rev. D} {\bf 41} 1796

\bibitem{R&T}
Reuter M and Tuiran E 2011 {\it Phys. Rev. D} {\bf 83} 044041

\bibitem{ligo}
See, for example, \url{https://en.wikipedia.org/wiki/List_of_gravitational_wave_observations} for a list of gravitational wave observations

\bibitem{eye}
Simpson A and Visser M 2022 {\it JCAP} {\bf 03} 011 (arXiv:2111.12329 [gr-qc])

\bibitem{S&S}
Smailagic A and Spallucci E 2010 {\it Phys. Lett. B} {\bf 688} 82 (arxiv:1003.3918 [hep-th])

\bibitem{S&S0}
Stuchl\'{\i}k Z and Schee J 2019 {\it Eur. Phys, J} {\bf 79} 44

\bibitem{take}
Takeuchi S 2016 {\it Class.Quant.Grav.} {\bf 33} 225016 (arXiv:1603.04159 [gr-qc])

\bibitem{TorresVoids}
Torres R 2013 {\it Class. Quantum Grav.} {\bf 30} 065014

\bibitem{TorresIns}
Torres R 2013 {\it Phys. Lett. B} {\bf 724} 338 (arXiv:1309.1083 [gr-qc])

\bibitem{dust2014}
Torres R 2014 {\it Phys. Lett. B} {\bf 733} 21

\bibitem{TorresExt}
Torres R 2017 {\it Gen. Rel. and Grav.} {\bf 49} 74 (arXiv:1702.03567 [gr-qc])

\bibitem{TorresReg}
Torres R and Fayos F 2017 {\it Gen. Rel. and Grav.} {\bf 49} 2 	(arXiv:1611.03654 [gr-qc])

\bibitem{Tosh}
Toshmatov B, Ahmedov B, Abdujabbarov A and Stuchlik Z 2014 {\it Phys. Rev. D} {\bf 89} 104017 (arXiv:1404.6443 [gr-qc])

\bibitem{Tsuka}
Tsukamoto N 2018 {\it Phys. Rev. D} {\bf 97} 064021 (arXiv:1708.07427 [gr-qc])

\bibitem{Wald}
Wald RM 1984 {\it General Relativity} Chicago : University of Chicago Press

\bibitem{Weinberg}
Weinberg S 1972 {\it Gravitation and Cosmology: Principles and
applications of the general theory of relativity} New York: John Wiley \&
Sons.

\bibitem{Y&Y}
Yagi K, Yunes N and Tanaka T 2012 {\it Phys. Rev. D} {\bf 86} 044037 (arXiv:1206.6130 [gr-qc]

\bibitem{ZM}
Zakhary E and McIntosh C B G 1997 {\it Gen. Rel. and Grav.} {\bf 29} 539

\end{thebibliography}
\end{document}